# Model-Based Trace-Checking

Yvonne Howard, Stefan Gruner, Andrew M Gravell, Carla Ferreira, Juan Carlos Augusto
DSSE, Department of Electronics and Computer Science, University of Southampton
Southampton, SO17 1BJ
Email: ymh@ecs.soton.ac.uk

**Abstract**

*Trace analysis can be a useful way to discover problems in a program under test. Rather than writing a special purpose trace analysis tool, this paper proposes that traces can usefully be analysed by checking them against a formal model using a standard model-checker or else an animator for executable specifications. These techniques are illustrated using a Travel Agent case study implemented in J2EE. We added trace beans to this code that write trace information to a database. The traces are then extracted and converted into a form suitable for analysis by Spin, a popular model-checker, and Pro-B, a model-checker and animator for the B notation. This illustrates the technique, and also the fact that such a system can have a variety of models, in different notations, that capture different features. These experiments have demonstrated that model-based trace-checking is feasible. Future work is focussed on scaling up the approach to larger systems by increasing the level of automation.*

## 1   Introduction

From the tester's perspective, tracing is perhaps considered a last resort. When a program or system crashes, it may be necessary to analyse a trace recorded in a log file which can be tedious. A trace viewer [Helmbold90] can help with this task, but human intuition is still required. An alternative approach, proposed in this paper, is to use an executable formal model of the program or system under test to automate the analysis of traces in support of testing. Trace code can be added semi-automatically to the program under test, so that any execution of the system (whether for testing purposes, or in deployment) is logged. Such logs can be played back through a standard model-checker or a special purpose trace analysis program. Erroneous executions will then be detected, even if they did not lead to any obvious and visible system failure or outage.

Note that as disk space is becoming cheaper, so larger trace files can feasibly be kept (and indeed, for security reasons, are often *required* to be kept). Moreover system kernels are becoming more reliable. If a problem arises, an exception handler will be called, which can be used to force output to be flushed to the disk. This makes it possible to buffer trace output, sometimes called lazy writing, which can be significantly more efficient.

Specification-based testing is not new [Richardson 89, Bochmann89]. Test oracles were generated from formal annotations (assertions) or from semi-formal specifications of communication protocols. Trace checking in distributed systems has also been studied [Jard94] including the problem of obtaining coherent global snapshots from local state. In this work the specifications were conjunctions of local predicates, and thus restricted to a certain class of safety properties.

Here we propose the combination of testing, tracing, executable formal models, and automated analysis. Andrews [Andrews98, Andrews00] describes a similar approach, but based on finite state specifications. In our work, the formal model can be finite state, as in traditional model-checking, or potentially infinite state, as in popular formal specification methods and notations such as B, VDM, and Z.

From the formal methods perspective, it has been a long standing problem to interest a wider section of the development community. Though it has been accepted for a number of years

[MoD91] that safety critical systems require rigorous development up to and including in some cases formal specification and verification, the high up-front costs of these approaches have largely precluded their use in non-critical applications. It has been argued [Wing90] that writing a formal specification confers benefits without needing to incur the cost of formal verification. The benefits claimed include improved understanding of the system, leading to fewer requirements capture and high level design errors, which in turn promise to reduce overall development costs. There is however little evidence to back up the notion of reduced development cost (IBM quote fewer errors in CICS development [Phillips90], which implies lower cost, but this is only one example).

Recently, model-checking [ClarkeES99] has proved successful in finding errors, particularly in hardware, network protocols and embedded software control systems such as railway signalling. The emphasis here is on detecting errors, not on proving their absence. Modelling tools such as Spin and ProB [Holzmann97, Leuschel 03] enable models to be written in days, used in model-checking, and then discarded [AugustoLBF03]. Different models can be created to explore different aspects of the systems behaviour. It is even possible to use a variety of model-checkers, for example Alloy [Jackson02] for data modelling, and Spin [Holzmann97] for control flow and concurrency checking. This means, in particular, that it is not necessary to capture all important behaviours in a single model (unlike with traditional formal specification).

This rather more pragmatic application of formal methods has the potential for significantly increasing their adoption in other areas of software development. Unfortunately, practical model-checking is currently limited to finite state systems – infinite state model-checking is still a research problem.

Other approaches include the animation of formal models [Fuchs92], which are considered controversial by some [Jones89], but a more fundamental concern is that the cost benefit of animation is probably not significant. Potentially more helpful is the generation of test cases from formal models [Dick93] but it has yet to be determined that such techniques (which rely to some extent on theorem proving techniques such as automatic simplification of expressions) can scale up to larger systems.

Our novel idea is therefore to
a) implement the program or system to test, using any suitable development method (it is not necessary to start from a formal specification, though this may help in the next step),
b) apply tests in the usual variety of ways (manual, random, or systematically designed test scripts),
c) before executing the tests, add (preferably automatically) trace code that captures each significant operation and event, in order to capture a significant volume of trace data,
d) write a formal model, or number of models, in a convenient notation, one for which suitable tools exist,
e) take the generated traces and run them through general or special purpose tools to check the traces against these models.

Steps a) and b) imply that trace-checking can easily be added to any existing development process. This is a very *low cost* or lightweight application of formal methods.

Aspect oriented [Walker98] programming tools such as AspectJ [Kiczales01] simplify the automated addition of trace code as in step c).

With respect to step d), the advantage of writing more than one model is that it may be easier this way to capture each interesting feature of the system under test. Writing a single specification (in a single notation) to cover all features can be a significant challenge.

Step e) means the models must be "executable" only insofar as they can be used to *test* in an efficient manner that a given trace is valid – *generating* all possible traces is not required. Note that in the case study described below, we use only existing general purpose tools (model-checkers and animators) for trace-checking. This is cheaper than developing a special purpose tool, since all that is required is to import data generated by, for example, Java into a tool programmed in, say, C or Prolog.

In this paper, section two describes our case study and how we added tracing components to a web application. Section 3 describes checking traces from the web application in a ProB animation of a B model, similarly, in the fourth section, we describe trace-checking in a SPIN model specified in Promela. Section five presents some of our conclusions and intentions for future work.

## 2 Case Study

Our case study is a travel agency that offers travel booking services to its customers via the Internet. A customer logs on to the travel agency welcome page, chooses the booking service they require (a hotel room reservation, or a rental car etc) and offers the credit card they will use to pay for the service when booked. The travel agency contacts appropriate suppliers and attempts to book the customers requirements, reporting back to the customer what has been booked on his behalf.

This case study is interesting for a number of reasons; it has components (travel agents, hotels and car rentals) that are distributed over the internet. There are many possible instantiations of the network of components. Each of the participating components has its own version of the booking data and yet the versions need to be consistent. And lastly, the case study is an e-business application and therefore uses the services of a complex layer of 'middleware'; web servers, object databases and java component technology through offered interfaces.

A typical instantiation has a number of travel agents, hotels, car rentals and customers. From the customer's perspective, they only interact with the travel agent; the suppliers of their hotel rooms or rental cars are hidden. From the supplier perspective, they only see the travel agent; they have no direct contact with a customer. From the travel agent's perspective they can see their customers and their suppliers; however bookings made with suppliers by other travel agents are hidden. The distributed nature of the travel agency and the complexity of the possible interactions mean that it is not easy to test.

There are business critical features of the system; we are asking for a credit card payment, we want to be sure that a reservation for a customer recorded by the travel agent is also recorded by the supplier of the service, and we want to preserve commercial privacy of the information.

We implemented the case study as a web application for Apache Tomcat web servers using Java Server Pages (JSP's), Java Servlets and Java Database Connectivity (JDBC). Booking data for the travel agent and supplier components was stored in Microsoft SQLServer2000 DBMS via JDBC, which also provided a 'name discovery' service for the interacting components to locate the suppliers of services they need.

By modelling the implementation, rather than a specification, we capture the behaviour of the implementation and can use the model to check that behaviour. If we want to use a model and model checker to explore the behaviour of our implementation, we should be sure that the models faithfully represent some abstraction of the implementation. Directing the modelling to aspects of the behaviour of the implementation and instrumenting our implementation to produce traces of tests, we were able to use the model checkers to test whether those traces violated any of the desired properties.

Interesting properties of the system that can be tested by checking the traces through a model checker include:
> 'if a car is requested and there is a car available one is eventually booked'
> 'no request goes ahead without Credit Card approval'

In this paper we will show ProB checking test traces through a B model of the implementation and SPIN checking test traces through a model written in Promela. [Holzmann97]

## 2.1 Adding Tracing
The next step in our approach is to add traces to the implementation. We used JSP beans to capture and record information about the implementation. Using bean technology means that we have a component with an interface structure already built into the web technology we are using. At each point that we make a trace record, we uses two components; a bean which makes a trace object, collecting state at the point of insertion, and a bean that writes the information to an SQL DBMS via JDBC. In our case study, all components of the travel agency, wherever located on the internet, write traces to the same SQL database

### 2.1.1 Trace components: tracebean
The trace bean class is a simple class with setter and getter methods for each attribute. Deploying the component is simple, when the trace is inserted in a jsp, the *jsp: usebean* interface manages the trace object creation, re-use and subsequent destruction. The same trace bean class is used in the java servlets but object creation and destruction has to be managed by the implementer.

### 2.1.2 Trace components: traceSqlbean
The trace SQL bean class provides a connection to the remote database via the JDBC interface and creates an object to write the trace bean object into the database. Similarly with the trace bean component, we can take advantage of JSP and servlet technology for deployment

Very often trace activity is logged into a simple text file but we have chosen to use an SQL database to collect the traces for the following reasons:
1. we can collect sets of trace information that will satisfy the needs of different members of the testing community; model checkers testing behaviour, testers carrying out traditional testing
2. we can easily filter the traces using SQL to provide formats that can be automatically fed into model checking tools
3. we can provide a dataset to check the trace for a particular property
4. we can be extravagant with the amount of data that we collect because we can use the database to manage the storage.
5. we can use the internet to connect all of the travel agency components to a remote SQL database via JDBC and therefore capture the trace of activity of distributed components

Having established a component structure for traces, the view of what we should collect was decided from the needs of the testing community for planned investigations and the tools they will use. Because we are using beans, changing the attributes of the bean objects to accommodate the needs of the testers as they explored the problem domain was reasonably simple. As far as possible we collected data that was available from the implementation either from user input or from interrogating the request and session objects. For example the traceComponent (the implementation component active when the trace is written) is obtained from the request object URI attribute.

At any point in our implementation, a trace is inserted with the following code (JSP example)

```
<jsp:useBean id="trace" class="tracer.TraceBean" scope="page"/>
<jsp:useBean id="traceSql" class="tracer.TraceSqlBean" scope="page"/>
…
//make a trace bean
   String bop = "enterCard" + "(" + session.getId()  + ")";
   String tId = session.getId();
   trace.setTraceId(tId);
   trace.setSId(session.getId());
   trace.setTraceUID(session.getAttribute("Thecustomer").toString());
   trace.setTraceBooktype(session.getAttribute("Thebooktype").toString());
   trace.setTraceCcType(session.getAttribute("ccType").toString());
   trace.setComponent(request.getRequestURI().toString());
   trace.setTraceBopName(bop);
//add it to the SQL database
   traceSql.addTrace( trace );
```

### 2.1.3   Where should we add a trace bean?

It would seem that we should add a trace when we capture new information, change component, and change state. Feedback from the models informs our choice of which state changes are significant and where a trace record should be made. For example: If we consider providing traces that will be checked against a B model; when the customer visits the welcome page and logs in, we capture their name. This corresponds to the B operation 'login' which takes the customer's name as a parameter. The 'log in' component of the implementation is instrumented to add a trace record to the trace database at this point.

In our case study, we inserted the tracing components manually, but the process could be automated and this is the subject of future work.

The point of correspondence between the implementation and the models is, at the moment, a matter of manual inspection. This, understandably, is a more error prone activity than if the attribute could be generated automatically from data already captured in the implementation. Close co-operation of the modellers and implementers to agree on naming will help to reduce the errors; for example if B model operations are designed to have the same names as the travel agency classes, servlets and JSP's. However if there is more than one B operation associated with an implementation component, then it will require judgement to determine at what point an equivalent B operation would be enabled. The first "error" we discovered using this method was in fact an artefact which arose because the trace captured data at the wrong point in the code.

One difficulty we encountered was the possibility that we could repeat a B operation name in two sequential trace records (where for example, a component has changed but the B operation has not changed.) We would prefer not to lose the distinction between components, but the ProB model checker requires that the operation name isn't duplicated. However because we are using an SQL database we can use DISTINCT to produce a trace dataset of B operations without duplications for ProB whilst retaining the richness of information for the other purposes.

In order to convert traces into a form suitable for model-checking, it is likely that data will have to be compressed into a "finite state" form. For example, our system tracks interactions through the system using session IDs. These are large random numbers generated by the web server. In any given trace, only a small number of session IDs are actually present, which we have to recognise and convert into a finite state type such as *Session* ≡ {*s*1, *s*2, *s*3, *s*4}.

## 3  Modelling our Travel Agency in B

B is a model-oriented formal notation and is part of the B-method developed by Abrial [Abrial96]. In the B-method, a system is defined as an abstract machine consisting of some state and some operations acting on the state. The invariant provides the logical properties that must be preserved by the system.

For the travel agency we have defined an abstract machine where the state contains information about user sessions, hotel and car rental bookings, and most operations replicate the interface presented to the user by the travel agency, like login and enterCard. Besides those the machine has operations that perform in a single step the service requested by the user (like bookRoom and unbookCar), which abstracts from the complexity of the implementation that requires several operations to implement a request.

The abstract machine has several invariant clauses; each of the clauses describes a property that must be preserved by the travel agency at all times. Next, we describe informally each one of the six invariant clauses:
1. This clause states that if a user has a booking, then it must have an assigned hotel where all the bookings have to be done. Conversely, if a user has an assigned hotel, then they must have a booking.
2. The second invariant clause states the same property, but for the car rentals.
3. This invariant clause says that all hotel bookings of a user must be done in the hotel assigned to that user.
4. The fourth clause states the same property, but for the car rentals bookings.
5. This invariant clause says that if a session does not have a valid card, the travel agency will not perform any service (booking or unbooking).
6. This clause states that if a session has a valid card, the travel agency will attempt to provide the requested service.

A useful property that does not appear in the invariant is the one that states that:
- if a user wants to book a room and there are available rooms, a room will be booked for that user

Modelling this property in B raises problems, as it contains a temporal ordering of events: if in t1 there is a room available, then in t2 one of the available rooms will be booked. To overcome this problem we modelled instead a similar property:
- if the request for booking a room was not fulfilled, then there must not be any rooms available.

### 3.1  Comparing Execution Traces using ProB

We assume at this point that the implementation and B-model are in good correspondence by virtue of traditional quality assurance methods like code inspection.

Changes in state in the B machine are caused by B operations. By manual inspection, we inserted trace beans in our instrumented version of the implementation at points where the change in state of our implementation corresponded to changes in state in the B machine (as described in section 2.1.3). This means that we capture a trace in our implementation at the same point that a B operation will be enabled in the B machine.

Originally designed for theorem proving rather than model checking and animation, the B language and method on its own cannot provide the trace checking for our higher level testing approach.

ProB [Leuschel01], [Leuschel03] is a new tool that provides animation, visualisation and model-checking for B machines. It is implemented in SICStus Prolog, exploiting advanced logic programming features such as co-routining and constraint solving. This gives performance comparable to more mature tools such as FDR and Spin. The input for ProB is the XML encoding provided by Bruno Tatibouet's jbtools B-to-XML parser. Animation (or interactive state exploration) uses a TCL/TK interface, and the dot open source graph drawing software. Model-checking (or exhaustive state exploration) can be used to check, for example, that a machine's invariant is not violated. ProB is currently still under active development. Improvements planned include the generation of test-cases from B machines, and model-checking for temporal logic properties and infinite state systems.

Our model–based trace-checking testing procedure works as follows:

1) Execute a test run of the implementation to record a test trace, either autoamatically or by manual testing.
2) Select the relevant trace data from the database using SQL to provide input for a ProB animation of the B model.
3) Load the model B machine and the trace into ProB and allow the animator to re-play the steps of implementation trace in terms of the currently available operations in the ProB animation of the B machine.
4) The higher level test is *passed*, if re-play (3) was possible. The higher level test *failed* if there was no combination of available B-machine operations in (3) could reproduce the test run described by the implementation trace.

An example of a set of trace data selected from the trace database and the subsequent history of its animation in ProB follows. Note that the name of the B machine and its initialisation has been added.

Customer user1 visits the welcome page and logs into the system. User1 decides to book a hotel room and is asked to enter some credit card data before the agency will process his request. The credit card information given by user1 is wrong, so the travel agency rejects their request and redirects him back to the point where a service can be chosen. In the meantime, user1 has changed his mind and decides to cancel a hotel room booking. Asked for the credit card details again, user1 gives a proper answer this time and the travel agent responds that their request failed. Given the choice between another booking and logging out, user1 decides to try to cancel a hotel booking again (perhaps he didn't believe the travel agent that he hadn't any rooms booked), receives the same response and this time decides to logout and leaves the travel agency. This little experiment leaves the following trace in our database:

```
machine('TravelAgency').
    initialise_machine.

    login(user1).
    choice(ss1).
    chooseService(ss1,H).
    enterCard(ss1).
    redoCard(ss1).
    choice(ss1).
    chooseService(ss1,U).
    enterCard(ss1).
    pickShop(ss1).
    '-->'(respUnbookCar(ss1),_).
    choice(ss1).
    chooseService(ss1,U).
    enterCard(ss1).
```

```
    pickShop(ss1).
    '-->'(respUnbookCar(ss1),_).
    logout(ss1).
```

This is a subset of the trace data from an execution of the implementation, the whole trace is richer, but only this subset is required for B and ProB, and matches exactly the input requirements for each B operation and outputs from a B operation. The first part of each record is the name of the B operation, the part in brackets is the input parameters or outputs. The parameters are derived from captured trace data. For example:

```
    chooseService(ss1,H).
```

The B operation 'chooseService' takes the session number (ss1) and booking service (H) as parameters. 'H' is directly derived from other trace data and indicates that the user has chosen to make a hotel booking, but the session numbers automatically assigned to the different web service sessions that track a customer's transaction, at 32 characters long, are cumbersome for animation so we use a translation program to provide a B operation parameter with the same format as its initialised set of sessions.

This subtrace can indeed be re-played in ProB in the initialised B machine where all the necessary operations are available when required – and thus the test is regarded as passed. The ProB operation history is as follows, showing the last activity first:

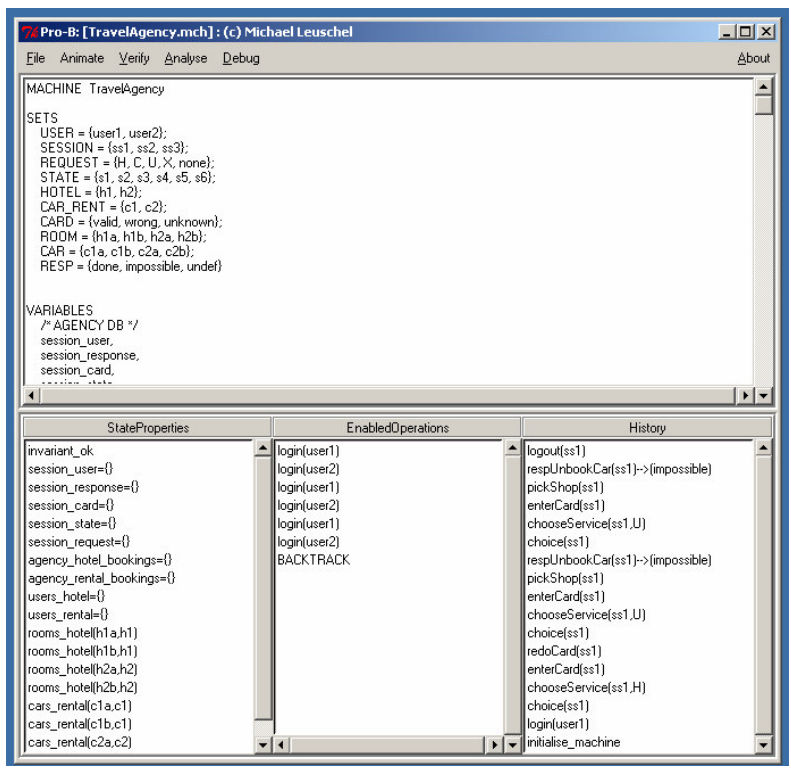

This figure shows the ProB animator and model checker. The initialised B model is shown in the top window; the bottom part is divided into three: the first showing the model checker computation of the state properties, the middle window shows the enabled operations at each state and the right-hand window shows the history of sequential operation steps followed. In trace-checking, the trace provides the next operation to be selected from the enabled operation. In the figure shown, the last operation was 'logout(user1)', and the only enabled operations are to 'login' one of the set of users.

Ongoing and future work is aimed at improving the level of automation of our test method such that long traces resulting from massive test avalanches can be processed and checked within an acceptable amount of time.

In checking such a trace, one problem that arises is that of specification non-determinism. We have captured only the input/output behaviour of the system, not internal choices which are abstracted in a typical specification. Our Prolog-based trace-checker handles this through back-tracking execution. If the specification is highly non-deterministic, however, the amount of back-tracking could become problematic, and slow down execution of the trace-checker. To reduce this overhead, or avoid it altogether, we have captured sufficient details of the internal behaviour of the system so that any non-deterministic choice is captured as soon as it is made. This requires the formal specification to capture not only the external behaviour, but also, at least to some extent, its internal design decomposition.

## 4 Using *SPIN* to Analyse Traces

Model checking [ClarkeES86] can be used to check whether a logical property holds for a finite-state system. A particularly successful implementation of this approach is *SPIN*, [Holzmann97] that has been widely accepted as a tool to support the verification stages in software development. It allows a series of useful verification actions in relation to a system specification with emphasis on efficiency. *SPIN* offers the possibility of performing simulations and verifications. Through these two modalities the verifier can detect absence of deadlocks and unexecutable code, to check correctness of system invariants, to find non-progress executions cycles and to verify correctness properties expressed in propositional linear temporal logic (*PLTL*) formulae. *Promela* is the specification language of *SPIN*. It is a C-like language enriched with a set of primitives allowing the creation and synchronization of processes, including the possibility to use both synchronous and asynchronous communication channels. We refer the reader to the extensive literature about the subject as well as the documentation of the system at Bell Labs web site (http://netlib.bell-labs.com/netlib/spin/whatispin.html); from now on we will assume some degree of familiarity with this framework.

In our case study we made use of *SPIN* in several ways. We focus in this section on two different ways of using this model checker to process traces produced by the prototype. One approach we followed was rewriting the trace in *Promela* and then checking the trace was consistent with an associated *PLTL* formula, see section "Using Temporal Logic Formulas as Partial Specifications" below. We also use trace sequences to build a *PLTL* formula that allowed us to check if equivalent steps can be made by *SPIN* executing a *Promela* specification., see section "Using Traces as Temporal Logic Formulas" below.

Due to space constraints we cannot offer a complete Promela model for our case study and we have to restrict ourselves to providing below a very brief description of it. The complete model, fully documented, can be seen as an appendix to [AugustoFGLN03]. In principle the different interacting parts of the system are reflected in the Promela model as processes, i.e., there are processes representing: the user, the travel agent, and each one of the shops. Other processes were introduced to match the conceptual division of classes in the prototype, making it easier to associate faulty parts of the model and the prototype. Communication between the user, the travel agent, and the shops is modelled via synchronous channels. Channels are used to pass requests from the user to the travel agent and then to some shop, or to get feedback from the shops about if the operation was or was not successful. Operations are registered by shops and the travel agent in structures that mimic the databases implemented in the prototype by using JDBC technology.

## 4.1 Using Temporal Logic Formulas as Partial Specifications

One way to check consistency between the traces obtained from the prototype with the basic properties of the system is by first transforming the traces into *Promela* code, i.e., a sequence of steps that mimic those taken in the prototype and then checking they are consistent with a related *PLTL* formula. In this case the *PLTL* formula can be seen as a simplified specification of an aspect of the system.

*Example.* Let us suppose we consider the problem of asking for a resource (e.g. a room from a hotel) for the second time when the first booking provoked exhaustion of the resource in the provider shop. Because the reservation strategy is incomplete, the system will only book subsequent resources from the same provider as the initial booking. After the initial booking exhausts the available resources, the travel agent will not attempt to book with another provider, even if those other providers have availability.

The traces considered below focus on a sequence of two steps. The first step is a successful attempt to book a room in Hotel1 that as a side effect provokes the Hotel1 to be full. The second attempt to book a room will be addressed to the same hotel and will fail.

The formula to be checked is: [] ((<>(requested && available)w) ==> <> allocate), more informally: "always when a resource is requested and there is one available, then one should be eventually allocated". Here "requested" means ((BookType=0) or (BookType=1)), "available" means ((RoomsAvailableHotel1>0) and (RoomsAvailableHotel2>0)) and "allocated" means that a room has been effectively booked (allocated=true). These definitions can be used to build the *PLTL* formula to be checked with *SPIN*. The specification below contains a trace with one booking and we can check with *SPIN* the sequence of states visited in the trace is consistent with the formula.

```
byte UserID,            /* a number identifying a user. e.g. between 1 and 10  */
    BookType,           /* the choice made bythe user (0:"car", 1:"hotel", etc) */
    CCType,             /* a credit card brand. (0:"VISA",  1:"MC", 2:"WRONG")  */
    SupplierName,       /* identifying a shop (1:"Hotel1", 2:"Hotel2", etc)     */
    RoomsAvailableHotel1, /* the number of available rooms in Hotel1 (0...)     */
    RoomsAvailableHotel2, /* the number of available rooms in Hotel2 (0...)     */
    RoomBooked,         /* a number identifying a room, e.g., between 1 and 5   */
    ShopAnswer;         /* the shop can inform the TA about the status of the
                           operation, e.g., 0:"impossible", 1:"done", etc       */
bool requested, available, allocated;
/* boolean variables used to identify states of interest during the trace */

init{ /* first login */
    UserID=2; BookType=1;
    requested=true;
    CCType=1; SupplierName=1; RoomsAvailableHotel1=1; RoomsAvailableHotel2=1;
    available=true;
    RoomBooked=1; ShopAnswer=1;
    allocated=true;}
```

But, if we add to the above specification the rest of the trace registering that a second booking is requested and the situation is such that 1) a second booking is requested also to Hotel1 when it is full and 2) Hotel2 has a room available:

```
/* second login */
UserID=2; BookType=1;
requested=true;
CCType=1; SupplierName=1; RoomsAvailableHotel1=0; RoomsAvailableHotel2=1;
available=true; ShopAnswer=0;
allocated=false
```

*SPIN* will prove that the enlarged sequence of steps does not validate the formula, i.e. will detect that despite there being a room available in Hotel2, the system will not allocate that room to the user as it reaches the state "allocated=false".
*(End Example)*

## 4.2 Using Traces as Temporal Logic Formulas

We may also benefit from exploring as much as possible the extent to which model is consistent with the implementation. We can use information produced by traces to verify that we obtain consistent behaviour when we explore the corresponding paths in the *Promela* model. We have several options that will accomplish this.
- we can use the results witnessed during implementation testing
- we can use interactive (user guided) simulation
- we can transform the testing conditions into a formula written in Linear Temporal Logic (LTL) [Holzmann97] and check its validity.

We have explored a method that combines these options in an automated or semi automated manner. We can transform a sequence of actions taken while testing the prototype into a temporal formula and then by using *SPIN* check that an equivalent sequence of actions can be made in the model. In order to automate the link between the output traces from the implementation and the input traces required by the model checker, we created a file that allows us to translate automatically implementation trace variable names to names used in the model. For example;

*corresponds([cctype,mc], [[ccbit1,1], [ccbit2,0]])*
*corresponds([cctype,wrong], [[ccbit1,1], [ccbit2,1]])*

where, the first line should be read as "if variable *cctype* has value *mc* in the implementation then variables *ccbit1* and *ccbit2* in the model have values 1 and 0 respectively".

*Example.* We carried out a test to determine whether it was possible to make the same number of credit card inputs as in the implementation.
a) Firstly, we manually tested in the implementation that we can enter first three "wrong" credit card brands before entering a valid one, "mc", and
b) Then we used a program to build an *LTL* formula that allows us to check if that is a possible scenario in the model. The resulting file, "formula.ltl", will have the following content:

```
#define p1 (ccbit1==1 && ccbit2==1)
#define p2 (ccbit1==1 && ccbit2==1)
#define p3 (ccbit1==1 && ccbit2==1)
#define p4 (ccbit1==1 && ccbit2==0)
/*
* Formula As Typed:    <>(p1 && (<>p2 && (<>p3 && (<>p4))))
*/
```

c) We used the formula in the *LTL Property Manager* section to generate the "never claim" (by pushing just two buttons). Using SPIN's No Executions (error behavior)"option to force a counterexample that confirms the sequence is possible.

*(End Example)*

The main problem faced whilst exercising the interaction between the prototype and *SPIN* is the synchronization of output/input traces. Filters and translators can be written to help in making these steps as automatic as possible but, as discussed earlier, close co-operation of modellers and implementers in naming reduces the amount and complexity of translation.

## 5   Conclusions and future work

In this work we have
- added tracing code, as trace beans, to our JSP implementation
- collected traces in a database, by exercising the implementation both interactively and randomly
- extracted traces and converted them into a format suitable for use with pre-existing and standard model-checking and animation tools
- verified these traces using ProB and Spin against models written in B and Promela, and also against temporal logic properties expressed in LTL.

So far, these steps have all been manual. Each of them, however, is very mechanistic and it is our intention to automate them in the near future.

Improvements to ProB are planned, in particular monitoring which cases in each operation have been executed, and how often, giving a measure of partition coverage. (At present, ProB only reports the number of states that have been reached, and whether its state exploration has reached all possible states.)

Compared with, say, introducing executable assertions into the program under test, this approach has the following advantages
1. formal models, for example in temporal logic, can easily capture *liveness*, not just safety properties
2. the formal model is interpreted by a pre-existing standard tool – avoiding the need to encode, say, the B notation in Java
3. it is possible to check existing traces for properties that had not been thought of at the time the trace was captured
4. the traces can be checked against a range of models, in a range of notations, using a range of tools

**Acknowledgements**
This work forms part of the ABCD project, which is funded by the EPSRC (GR/M91013/01), whose support we gratefully acknowledge.